\documentclass[preprint, nofootinbib]{revtex4-1}

\pdfoutput=1 
\usepackage{float}
\usepackage{caption}
\usepackage{textcomp}
\usepackage{amsmath}
\usepackage{amssymb}
\usepackage{graphicx}
\usepackage{subfigure}
\usepackage{diagbox}
\usepackage{color}
\usepackage{ulem}
\usepackage[bottom]{footmisc}
\usepackage{setspace}
\usepackage[unicode=true, bookmarks=false, breaklinks=false,pdfborder={0 0 1},colorlinks=false] {hyperref}



\begin{document}

\makeatletter

\@addtoreset{equation}{section}

\@addtoreset{equation}{subsection}

\makeatother
\pagestyle{plain}
\setcounter{page}{1}

\baselineskip16pt

\begin{titlepage}

\begin{flushright}

\end{flushright}
\vspace{8 mm}

\begin{center}

{\Large \bf S-duality transformation of $\mathcal{N}$ $=4$ SYM theory at the operator level   \\}

\end{center}

\vspace{7 mm}

\begin{center}

{\bf Shan Hu }

\vspace{5mm}
{\small \sl Department of Physics, Faculty of Physics and Electronic Sciences, Hubei University,} \\
\vspace{2mm}
{\small \sl  Wuhan 430062, P. R. China} \\

\vspace{5mm}

{\small \tt hushan@itp.ac.cn} \\

\end{center}

\vspace{8 mm}

\begin{abstract}

\end{abstract}

We consider the S-duality transformation of gauge invariant operators and states in $\mathcal{N}$ $=4$ SYM theory. The transformation is realized through an operator $ S $ which is the $ SL(2,Z) $ canonical transformation in loop space with the gauge invariant electric and the magnetic flux operators composing the canonical variables. Based on $ S $, S-duals for all of the physical operators and states can be defined. The criterion for the theory to be S-duality invariant is that the superconformal charges and their S-duals differ by a $ U(1)_{Y} $ phase. The verification can be done by checking the S transformation for supersymmetry and special supersymmetry variations of the loop operators. The fact that supercharges preserved by BPS Wilson operators and the S-dual BPS 't Hooft operators differ by a $4d  $ chiral rotation could in some sense serve as a proof.

\vspace{1cm}
\begin{flushleft}

\end{flushleft}
\end{titlepage}
\newpage

\section{Introduction}

$\mathcal{N}$ $= 4$ super Yang-Mills (SYM) theory is believed to realize an $SL(2,Z)$ duality \cite{SL1, SL2}. The action of the duality group on the gauge coupling $\tau$ is
\begin{equation}\label{sd}
	\tau =\tau_{1} +i \tau_{2}=\frac{\theta}{2 \pi}+i\frac{ 4\pi }{g^{2}} \rightarrow \frac{a \tau+b}{c\tau +d}\;,\;\;\;\;\;\;\;\left(                
\begin{array}{cc}   
a & b\\ 
c & d \\  
\end{array}
\right)  \in SL(2,Z) \;.
\end{equation}
Theories with $\tau$'s related by the $SL(2,Z)$ transformations are physically equivalent.\footnote{Generically, S-duality makes $\mathcal{N}$ $= 4$ SYM theory with the gauge group $ G $ mapped into the one with the dual group $ ^{L}G $ \cite{sw1}. Here, we only consider $ U(N) $, which is self-dual.} Especially, when $\theta=0$, $a=d=0$, (\ref{sd}) reduces to $g \rightarrow 4 \pi /g$, the strong-weak duality of Montonen and Olive \cite{sw1, sw2, sw3}.

In Coulomb phase, the action of $SL(2,Z)$ on BPS states is well known \cite{SL1}. BPS mass spectrum is invariant under (\ref{sd}). The perturbative massive vector multiplets are mapped into the vector multiplets arising from the quantization of the monopoles or dyons. In conformal phase, when the theory exhibits the $ PSU(2,2|4) $ superconformal symmetry, the observables of interest are gauge invariant operators and their correlation functions. The $SL(2,Z)$ transformation will make a gauge invariant operator $ O $ mapped into a gauge invariant operator $ O' $. Correlation functions of $ O' $ in theory with the coupling constant $ \tau $ equal the correlation functions of $ O $ in the same theory with the coupling constant $(a \tau+b)/(c \tau+d)  $:
\begin{equation}\label{cor}
\langle \prod_{i} O'_{i}\rangle_{\tau}=\langle \prod_{i} O_{i}\rangle_{\frac{a \tau+b}{c \tau+d}}\;.
\end{equation}
Explicit conjectures have been made for the S-duality actions on the local operators \cite{op1, op3, op2}, line operators \cite{wh1, wh2}, surface operators \cite{14, 15} and domain walls \cite{16, 17}. Calculations of the correlation functions are made in \cite{op1, op3, wh3, wh4}.

$\mathcal{N}$ $= 4$ SYM theory is dual to the type IIB string theory on $ AdS_{5} \times S^{5}$ which is also $ SL(2,Z) $ invariant \cite{ac1, ac2, ac3}. The duality transformation of the latter offers clues for the duality transformation of the former. In type IIB, components of the supergravity multiplet are assigned with the definite $U(1)_{Y}$ charges, and the S-duality transformation can be seen as a $U(1)_{Y}$ transformation \cite{u1,u2,u3,u33}. The corresponding operators in SYM theory should have the opposite $U(1)_{Y}$ charges \cite{u4}, which then leads to the conjecture that correlation functions of $ 1/2 $ BPS operators in $\mathcal{N}$ $= 4$ SYM theory should exhibit the $U(1)_{Y}$ symmetry \cite{op1, op3}. In \cite{op2}, the action of the $SL(2,Z)  $ duality on the local gauge invariant operators is studied based on the conformal weight. It was suggested that for operator with the modular invariant conformal weight, such as the BPS operator, $ SL(2,Z) $ transformation will map it into itself up to a possible multiplicative factor if there is no degeneracy. Otherwise, the operator will transform as part of a finite or infinite dimensional $ SL(2,Z) $ multiplet, which is the situation for the Konishi operator.

In this paper, we will study the S-duality transformation for operators and states of $\mathcal{N}$ $= 4$ SYM theory in more detail. Based on the commutation relation between the Wilson and 't Hooft operators in canonical quantization formalism \cite{Hooft}, we propose a canonical commutation relation in loop space with the gauge invariant electric and magnetic fluxes composing the canonical variables. The canonical transformation in loop space is the $ SL(2,Z) $ duality transformation of the theory. The transformation is realized through an operator $ S $, based on which, the S-duals for all of gauge invariant operators and states can be defined. The criterion for the theory to be S-duality invariant is that superconformal charges and their S-duals differ by a $ U(1)_{Y} $ phase, or equivalently, supercharges and special supercharges transform with a $ U(1)_{Y} $ phase. The fact that supercharges preserved by BPS Wilson operators and those by BPS 't Hooft operators differ by a $4d  $ chiral transformation \cite{bps} would indicate the operator $ S $ making the Wilson operator transformed into the 't Hooft operator will also make the supercharges transform with a $ U(1)_{Y} $ phase thus could serve as a proof for the duality invariance.

With the criterion met, correlation functions of the dual operators are equal to the correlation functions of the original operators in a theory with the dual coupling constant as is in (\ref{cor}). The $ PSU(2,2|4)  $ irreducible state with the energy $ E(\tau) $ is mapped into a state in the same theory with the energy $ E(\frac{a \tau+b}{c \tau+d}) $ and the identical $ SO(3) \times SU(4) $ quantum number. Although the correlation functions of $ 1/2 $ BPS operators at distinct points are expected to transform with a $ U(1)_{Y} $ phase under the $ SL(2,Z) $ transformation according to AdS/CFT \cite{op1}, we show that not all of the $ 1/2 $ BPS operators themselves would transform in this way. How the two can be compatible with each other is also discussed.

The rest of the paper is organized as follows: in section 2, we review the S-duality transformation of the $ U(1) $ gauge theory; in section 3, we review the S-duality transformation in type IIB string theory; in section 4, we study the S-duality transformation of $\mathcal{N}$ $= 4$ SYM theory; the discussion is in section 5.

\section{S-duality transformation in $ U(1) $ gauge theory}

For Maxwell theory, S-duality transformation is implemented as  $ E_{i}\rightarrow B_{i} $, $ B_{i}\rightarrow -E_{i} $ both classically and quantum mechanically, $ i=1,2,3 $. In temporal gauge, the canonical operators are $ (A_{i},E^{i}) $, and the gauge invariant physical Hilbert space $ H_{ph} $ is obtained by imposing the Gauss constraint $ \partial_{i} E^{i}=0$ in $ H$. $ \partial_{i} B^{i}=0$ always holds but $ \partial_{i} E^{i}=0$ is only valid in $ H_{ph} $. Both $ E_{i} $ and $B_{i}   $ are physical operators. S-duality transformation can be realized via a unitary operator $ S $ in $ H_{ph} $:
\begin{equation}\label{201}
S\hat{E}_{i}S^{-1}=\hat{B}_{i} \;,\;\;\;\;\;\;\; S\hat{B}_{i}S^{-1} =-\hat{E}_{i} \;,
\end{equation}
where $ \hat{•}$ represents a projection in $ H_{ph} $. There is no unitary operator relating $ E_{i} $ and $ B_{i} $ in $ H $, since $\partial_{i} E^{i} =0 $ only in $ H_{ph} $. 

The situation for the $ U(1) $ Born-Infeld theory with the couping $ g^{2} $ is also similar \cite{1,2,3,980}. In temporal gauge, the canonical operators are $ (A_{i},D^{i}) $ and $ H_{ph} $ is obtained by imposing $ \partial_{i} D^{i}=0$ in $ H$. $ \partial_{0} A_{i}=E_{i}\neq D_{i}=\frac{\delta L}{\delta E^{i}} = D_{i}(g^{2}, A_{i},E_{i})$. $ D_{i} $ and $ B_{i} $ are physical operators. S-duality transformation is still realized via a unitary operator $ S $:
\begin{equation}\label{202}
S\hat{D}_{i}S^{-1}=\hat{B}_{i} \;,\;\;\;\;\;\;\; S\hat{B}_{i}S^{-1} =-\hat{D}_{i} \;.
\end{equation}
The Hamiltonian is 
\begin{equation}
h=\sqrt{1+g^{2}D_{i}D_{i}+\frac{1}{g^{2}} B_{i}B_{i}+(\epsilon_{ijk}D_{j}B_{k})(\epsilon_{imn}D_{m}B_{n})}-1\;,
\end{equation}
$ S\hat{h}(g^{2}) S^{-1}=\hat{h}(1/g^{2})$. Hamiltonians with the coupling constants $g^{2}  $ and $ 1/g^{2} $ are related by the unitary transformation $ S $.

Generically, for the arbitrary $ U(1) $ gauge theory in temporal gauge, the canonical operators are $ (A_{i},\Pi^{i}) $ with $ \partial_{i} \Pi^{i}=0 $ in $ H_{ph} $. The operator $ S $ is constructed as
\begin{equation}\label{che}
S\hat{\Pi}_{i}S^{-1}=\hat{B}_{i} \;,\;\;\;\;\;\;\; S\hat{B}_{i}S^{-1} =-\hat{\Pi}_{i} \;.
\end{equation}
In canonical quantization formalism, the definition of the S-duality transformation only depends on the field content and has nothing to do with the Hamiltonian. We can study the effect of the transformation on the Hamiltonian $ h(g^{2},B_{i},\Pi_{i}) $. If   
\begin{equation}
 S\hat{h}(g^{2},B_{i},\Pi_{i}) S^{-1}=\hat{h}(g^{2},-\Pi_{i},B_{i})=\hat{h}(\frac{1}{g^{2}},B_{i},\Pi_{i})\;,
   \end{equation}   
the theory is S-duality invariant. Otherwise, the original theory is dual to a different theory with the coupling constant $ \frac{1}{g^{2}} $ and the different Hamiltonian.

In (\ref{che}), $ S $ is defined through its action on the local gauge invariant operators in the ordinary space, but it could also be equivalently defined in loop space. For $ U(1) $ gauge theory, the Wilson operator is $ W(C)=e^{i w(C)} $ with the magnetic flux operator
\begin{equation}
w(C)= \oint_{C} A_{i} (x) dx^{i}   =  \int\int_{\Sigma_{C}} B_{i} (x) d\sigma^{i} \;. 
\end{equation}    
The projection of $ w(C) $ in $ H_{ph} $ is denoted as $\hat{w} (C)  $. The $ U(1) $ version of the 't Hooft operator is $ \hat{T}(C)=e^{i\hat{t}(C)} $ with the electric flux operator 
\begin{equation}
\hat{t}(C) =  \int\int_{\Sigma_{C}} \hat{\Pi}_{i} (x) d\sigma^{i}  \;.
\end{equation}   
The direct calculation shows \footnote{We make a rescaling here to make $ [A_{i} (x), \Pi_{j} (y)] =-2 \pi i \delta_{ij}\delta^{3} (x-y)$ since $ eg=2\pi $ according to the Dirac quantization condition.}
\begin{equation}\label{111q}
[\hat{t} (C'), \hat{w} (C)]=2 \pi i n \;,
\end{equation}
where $ C $ and $ C' $ are arbitrary two spatial loops at the equal time with the linking number $ n $. (\ref{111q}) could be taken as the canonical commutation relation in loop space and $(\hat{w}, \hat{t})  $ are canonical variables with the suitable constraints imposed to eliminate the unphysical configurations that could not be obtained from the spacetime gauge potential. From (\ref{111q}), the $ U(1) $ version of the commutation relation for Wilson and 't Hooft operators \cite{Hooft}
\begin{equation}
 \hat{W}(C) \hat{T}(C') = \hat{T}(C')\hat{W}(C) \exp (2 \pi i n)\;
\end{equation}
can be obtained.

From (\ref{che}), we have 
\begin{equation}\label{che1}
S\hat{t} (C)S^{-1}=\hat{w}  (C)  \;,\;\;\;\;\;\;\; S\hat{w}  (C) S^{-1} =-\hat{t} (C) \;.
\end{equation}
Conversely, if (\ref{che1}) is satisfied for the arbitrary spatial loops, (\ref{che}) will hold. (\ref{che}) and (\ref{che1}) are two equivalent ways to define $ S $. In nonabelian theory, the definition (\ref{che}) does not work since S-duality transformation can only be definitely determined for operators in $ H_{ph} $ but $B_{i}$ and $\Pi_{i}  $ are not gauge invariant any more. On the other hand, loop operators are always gauge invariant, so (\ref{che1}) is still valid. Later, we will try to extend (\ref{che1}) into the $\mathcal{N}$ $=4$ SYM theory to construct a S-duality transformation operator $ S $.

\section{S-duality transformation in type IIB string theory}

Type IIB string theory is $SL(2,Z)$ invariant. The 32 supercharges of Type IIB theory could be combined into two left handed Majorana-Weyl spinors $Q_{L}$ and $Q_{R}$,\footnote{Here, L and R refer to the chirality in string worldsheet.} satisfying both the Majorana condition
\begin{equation}
Q_{L}^{T}C=Q_{L}\;,\;\;\;\;\;\;\;\;Q_{R}^{T}C=Q_{R}\;,
\end{equation}
and the Weyl condition
\begin{equation}
\Gamma^{11}Q_{L} = Q_{L}\;,\;\;\;\;\;\;\;\; \Gamma^{11}Q_{R} = Q_{R}\;.
\end{equation}
It is convenient to combine two Majorana-Weyl spinors into complex Weyl spinors
\begin{equation}
Q^{\pm}=Q_{L}\pm iQ_{R}\;.
\end{equation}
S-duality transformation will induce a phase rotation to Weyl spinors
\begin{equation}
Q^{+}\rightarrow   (\frac{c \tau +d}{c \bar{\tau} +d})^{\frac{1}{4}}Q^{+}\;,\;\;\;\;\;\;\;\;Q^{-}\rightarrow    (\frac{c \tau +d}{c \bar{\tau} +d})^{-\frac{1}{4}}Q^{-}\;,
\end{equation}
where
\begin{equation}
\left(                
\begin{array}{cc}   
a    & b\\ 
c  & d \\  
\end{array}
\right)\in SL(2,Z)\;.
\end{equation}
$\tau$ is the complex scalar field built out of the dilaton and the the axion. Type IIB supergravity fields could be organized into a supermultiplet with the definite $U(1)_{Y}$ charge \cite{u1,u2,u3,u33}. The action of $ SL(2,Z) $ will make the fields with the $U(1)_{Y}$ charge $q$ transform as 
\begin{equation}\label{fieg}
F \rightarrow   (\frac{c \tau +d}{c \bar{\tau} +d})^{\frac{q}{4}}F\;.
\end{equation}

S-duality transformation makes F-string mapped into the $ (a,b) $ string, and the first quantization of them gives the string fields $ f^{(1,0)}_{m,n} $ and $ f^{(a,b)}_{m,n} $ that should also be mapped into each other. As the $ (m,n) $ excitation of the $ (a,b) $ string, $  f^{(a,b)}_{m,n} $ is non-perturbative in $ (1,0) $ F-string frame. However, the massless excitations for all of the $ (a,b) $ strings are the same supergravity multiplet related by a $ U(1)_{Y} $ rotation. In this sense, supergravity multiplet is universal in all of the S-frame.

\section{S-duality transformation in $ \mathcal{N} $ $ =4 $ SYM theory}

S-duality transformation in $ U(1) $ gauge theory is realized through a unitary operator $ S $ in physical Hilbert space. We will construct the similar $ S $ for $ \mathcal{N} $ $ =4 $ SYM theory with the gauge group $ U(N) $.

\subsection{Definition of the S-duality transformation}

The Lagrangian of the $ \mathcal{N} $ $ =4 $ SYM theory with the coupling constant $ \tau=\tau_{1}+ i\tau_{2} $ is 
\begin{eqnarray}\label{ll}
\nonumber L&=&\frac{\tau_{2}}{4 \pi}tr \{-\frac{1}{4}F_{\mu\nu}F^{\mu\nu}+\frac{\tau_{1}}{4\tau_{2}}F_{\mu\nu}\ast F^{\mu\nu}-i\bar{\Psi}^{a}\bar{\sigma}^{\mu}D_{\mu}\Psi_{a}-\frac{1}{2}D_{\mu}X^{I}D^{\mu}X^{I}\\&&
+\;\frac{1}{2}C^{ab}_{I} \Psi_{a} [X^{I},\Psi_{b}]+\frac{1}{2} \bar{C}_{Iab} \bar{\Psi}^{a} [X^{I},\bar{\Psi}^{b} ]+\frac{1}{4} [X^{I},X^{J}]^{2}\}\;.
\end{eqnarray}
$ D_{\mu}f=\partial_{\mu}f-[A_{\mu},f] $. In temporal gauge, $ A_{0} =0$, canonical fields are $(X^{I} ,\Psi_{a},A_{i} )$ with the conjugate momentum $ (\Pi_{I},\Pi^{a},\Pi^{i}) $. $ I=1,2,\cdots,6 $, $ a=1,2,3,4 $, $ i=1,2,3 $. Let 
\begin{equation}
\Omega^{A}= \partial^{i}\Pi_{i}^{A} -if^{A}_{BC} A^{i\;B}\Pi_{i}^{C}  +if^{A}_{BC}X^{I\;B}\Pi_{I}^{C}-if^{A}_{BC}\Psi^{a\;B} \Pi_{a}^{C}\;,
\end{equation}
$ A=1,2,\cdots,N^{2} $. The physical Hilbert space $ H_{ph} $ is composed by states satisfying $ \Omega^{A}\vert\psi\rangle=0 $. $ \Omega^{A} $ and $ H_{ph}  $ are $ \tau $-independent.

Hilbert space $ H $ could be decomposed into the direct sum of $ H_{ph}  $ and its complementary space: $  H =\bar{H}_{ph}  \oplus H_{ph}  $. A generic operator $ O $ could be correspondingly decomposed as 
\begin{equation}
O=\left(                
\begin{array}{cc}   
O_{11}    & O_{12} \\ 
O_{21}   & O_{22}  \\  
\end{array}
\right)\;.
\end{equation}  
$ \Omega^{A} $ and the physical operators take the form of 
\begin{equation}
\Omega^{A}=\left(                
\begin{array}{cc}   
\Omega^{A}_{11}    & 0 \\ 
0  & 0  \\  
\end{array}
\right)   \;\;\;\;\;\;\;\;\;\;\;O_{ph}=\left(                
\begin{array}{cc}   
O_{11}    & 0 \\ 
0   & \hat{O}_{ph}  \\  
\end{array}
\right)\;,
\end{equation} 
where $ \hat{O}_{ph}$ is the projection of $ O_{ph} $ in $  H_{ph} $.

Extending (\ref{che1}) into the $ U(N) $ theory requires the construction of the corresponding flux operators. For $ U(N) $ YM theory, the Wilson operator is
\begin{equation}
\hat{W}(C)=\frac{1}{N}\hat{tr} \left[  P \exp \{i \oint_{C} A_{i} (x) dx^{i} \}   \right] 
\end{equation}
in fundamental representation. $ \hat{tr} $ represents a projection in $ H_{ph} $. $ W(C) $ for the arbitrary spatial loops compose the complete gauge invariant observables of the theory. For the YM theory in temporal gauge, the fundamental field is $ A_{i} $. Polyakov has shown that the gauge equivalent class of $ A_{i}  $ can be extracted from the element of the holonomy group
\begin{equation}
\Phi(C,x(0))=  P \exp \{i \oint_{C} A_{i} (x) dx^{i} \}   \;,
\end{equation}
where all loops begin and end at the same fixed point $ x(0)  $ \cite{loop}. 
\begin{equation}
W(C)=\frac{1}{N}tr [\Phi(C,x(0))  ]\;.
\end{equation}
Suppose $ C^{n} $ is a loop beginning and ending at $ x(0)  $, winding $ C $ $ n $ times, then 
\begin{equation}
W(C^{n})=\frac{1}{N}tr [\Phi^{n}(C,x(0))  ]\;.
\end{equation}
With $W(C^{n})  $ for $ n=1,2,\cdots,N $ given, $\Phi(C,x(0))  $ can be determined up to a gauge transformation, so $W(C)  $ contains complete degrees of freedom of the theory. In $ \mathcal{N} $ $ =4 $ SYM theory, the complete gauge invariant observables should be the super Wilson operator built from the $ \mathcal{N} $ $ =4 $ vector multiplet. In \cite{su1, su2, su3}, super Wilson operators are constructed and could be taken as the Wilson operators for $ 10d $ YM theory in superspace.

With the super Wilson operator $ \hat{W}(C) $ given, the magnetic flux operator $ \hat{w} (C)  $ is obtained via
\begin{equation}
\hat{W}(C) =\exp \{i \hat{w} (C) \}
\end{equation}
following the definition in \cite{flux1}. $ \hat{w} (C) $ could act as the canonical coordinate in loop space with the conjugate momentum $\hat{t} (C)  $. The equal time canonical commutation relation is 
\begin{equation}\label{com}
[\hat{t} (C'), \hat{w} (C)]=\frac{2 \pi i n}{N}
\end{equation}
and 
\begin{equation}\label{com1}
[\hat{w} (C), \hat{w} (C')]=[\hat{t} (C), \hat{t} (C')]=0
\end{equation}
with $ n $ the linking number between $ C' $ and $ C $. For 
\begin{equation}
\hat{T}(C) =\exp \{i \hat{t}(C) \}\;,
\end{equation}
from (\ref{com}), 
\begin{equation}\label{AB}
\hat{W}(C) \hat{T}(C') = \hat{T}(C')\hat{W}(C) \exp (2 \pi i n/N)\;, 
\end{equation}
which is the commutation relation for the Wilson and 't Hooft operators \cite{Hooft}, so $ \hat{T}(C) $ can be taken as the 't Hooft operator.

Consider the canonical transformation in loop space:
\begin{equation}\label{abc1}
S[ \left(                
\begin{array}{cc}   
a & b\\ 
c & d \\  
\end{array}
\right) ;\tau]  \left(                
\begin{array}{c}   
\hat{w} (C) \\ 
\hat{t} (C)   \\  
\end{array}
\right)S^{-1}[ \left(                
\begin{array}{cc}   
a & b\\ 
c & d \\  
\end{array}
\right) ;\tau]   = \left(                
\begin{array}{cc}   
a & c\\ 
b & d \\  
\end{array}
\right)  \left(                
\begin{array}{c}   
\hat{w}(C) \\ 
\hat{t} (C)   \\  
\end{array}
\right)   \;.
\end{equation}
To preserve (\ref{com})-(\ref{com1}), $ ad-bc=1 $. Moreover, $(W(C) )^{N} =(T(C) )^{N}=1$, $ \hat{w}(C) $ and $ \hat{t}(C) $ are quantized in units $ 2 \pi/N $ \cite{flux1}, which also breaks $ SL(2,R) $ to $ SL(2,Z) $. The corresponding loop operators transform as
\begin{eqnarray}
\nonumber  S^{-1}\hat{W} (C)S&=&  \hat{T}^{-c}(C) \hat{W}^{d}(C) \\
\nonumber   S^{-1}\hat{T} (C)S&=&  \hat{T}^{a}(C) \hat{W}^{-b}(C)  \;.
\end{eqnarray}
This is the S-duality transformation rule for Wilson and 't Hooft operators in $ \mathcal{N} $ $ =4 $ SYM theory \cite{wh1, wt, tw}, so the canonical transformation operator $ S $ generates the $ SL(2,Z) $ duality transformation.

The unitary operator $ S $ in $  H_{ph} $ can also be extended into a unitary physical operator $ U$ in $  H  $:
\begin{equation}
U=\left(                
\begin{array}{cc}   
V    & 0\\ 
0  & S \\  
\end{array}
\right)\;,
\end{equation}
where $ V $ is an arbitrary unitary operator acting on $ \bar{H}_{ph}$. In this way, although not unique, the S-duality transformation of canonical fields can also be defined:
\begin{eqnarray}\label{u}
\nonumber && U^{-1} XU=\tilde{X}\;\;\;\;\;\;\;\;\;U^{-1} \Psi U=\tilde{\Psi}\;\;\;\;\;\;\;\;\;U^{-1} AU=\tilde{A}\\\nonumber &&
U^{-1} \Pi_{X}U=\tilde{\Pi}_{X}\;\;\;\;\;U^{-1} \Pi_{\Psi} U=\tilde{\Pi}_{\Psi}\;\;\;\;\;U^{-1} \Pi_{A} U=\tilde{\Pi}_{A}\;.\\
\end{eqnarray}  
(\ref{u}) could be regarded as a canonical transformation in ordinary space. In contrast to loop space, where $ S $ has a simple representation as an $ SL(2,Z) $ transformation, for the nonabelian theory in ordinary space, it is difficult to tell the exact relation between the S-dual fields and the original fields.

\subsection{Criterion for the S-duality invariance}

At this stage, the action of the S-duality transformation has been defined. The definition is based on the field content with no dynamical information involved. The dynamical information of $\mathcal{N}$ $=4$ SYM theory is encoded in superconformal charges.

For $\mathcal{N}$ $=4$ SYM theory with the coupling constant $\tau$, suppose $ G $ is a $ PSU(2,2| 4) $ generator with the $ U(1)_{Y} $ charge $ q $, 
\begin{equation}
G(\tau)= G[X, \Psi,A;\Pi_{X},\Pi_{\Psi},\Pi_{A}|\tau]\;.
\end{equation}
The generator constructed from the S-dual canonical fields with the coupling constant $ \frac{a \tau+b}{c \tau+d} $ is 
\begin{eqnarray}
\nonumber && \tilde{G}(\frac{a \tau+b}{c \tau+d})=  G[\tilde{X}, \tilde{\Psi},\tilde{A};\tilde{\Pi}_{X},\tilde{\Pi}_{\Psi},\tilde{\Pi}_{A}|\frac{a \tau+b}{c \tau+d}]\\ &=&
U^{-1}G(\frac{a \tau+b}{c \tau+d})
U=G'[X, \Psi,A;\Pi_{X},\Pi_{\Psi},\Pi_{A}|\tau]=G'(\tau)\;.
\end{eqnarray} 
The projection of $ G(\tau) $ and $G'(\tau)  $ in $ H_{ph} $ is $ \hat{G}(\tau) $ and $\hat{G}'(\tau)  $. Then the criterion for the theory to be S-duality invariant is \footnote{In (\ref{u1}), it is necessary to make a projection in $ H_{ph} $ to get $ \hat{G}  $ from $ G $. For example, when $ G $ is the supercharge $ Q $, since
\begin{equation}\label{hg}
\{ Q^{a}_{\alpha } ,Q^{b}_{\beta}\}=-2ig\int d^{3}x\;\epsilon_{\alpha\beta}tr\{\Omega X^{ab}\}
\;\;\;\;\;\;\;\;\;
\{ \bar{Q}_{a\dot{\alpha} } ,\bar{Q}_{b\dot{\beta} }\}=-2ig\int d^{3}x\;\epsilon_{\dot{\alpha}\dot{\beta}}\epsilon_{abcd}tr\{\Omega X^{cd}\}\;,
\end{equation}
there is no unitary transformation $ U $ making $ UQU^{-1} =e^{i\theta} Q$ and $ U\bar{Q}U^{-1}= e^{-i\theta} \bar{Q}$. On the other hand, in $ H_{ph} $, $\Omega=0$, so there can be a unitary transformation $ S$ making $ S\hat{Q}S^{-1}= e^{i\theta} \hat{Q}$ and $S\hat{\bar{Q}} S^{-1}= e^{-i\theta} \hat{\bar{Q}} $. } 
\begin{equation}\label{u1}
\hat{G'}(\tau)=\hat{\tilde{G}}(\frac{a \tau+b}{c \tau+d})=S^{-1} \hat{G}(\frac{a \tau+b}{c \tau+d} ) S= (\frac{c \tau +d}{c \bar{\tau} +d})^{-\frac{q}{4}}\hat{G}(\tau)\;.
\end{equation}
Especially, when $ G $ is the Hamiltonian, 
\begin{equation}\label{harmi}
\hat{H'}(\tau)=\hat{\tilde{H}}(\frac{a \tau+b}{c \tau+d})=S^{-1} \hat{H}(\frac{a \tau+b}{c \tau+d} )S=\hat{H}(\tau)\;.
\end{equation}
If (\ref{harmi}) is not satisfied, the original theory will be dual to another theory with a different Hamiltonian and the coupling constant $ \frac{a \tau+b}{c \tau+d}  $.

$\mathcal{N}$ $=4$ SYM theory is the low energy effective theory on $ D3 $ branes. To describe the physics on $ D3 $, we may select two sets of canonical fields $ (X,\Psi,A) $ and $ (\tilde{X},\tilde{\Psi},\tilde{A}) $ coming from the quantization of the open F-string and the open $ (a,b) $-string, and then construct two theories with the coupling constants $ \tau $ and $ \frac{a \tau+b}{c \tau+d} $.\footnote{In Coulomb phase, the perturbative degrees of freedom are vector multiplet, from which, the Hilbert space is constructed. The configuration space also contains monopoles and dyons, and the quantization of them gives a set of vector multiplets labelled by $ (a,b) $ in the same Hilbert space. The action of $ SL(2,Z) $ is also defined as a transformation from one set of the vector multiplet into another.} Two theories have the same physical Hilbert space $ H_{ph} $ with the superconformal charges related via (\ref{u1}). The dynamics of $ D3 $ takes the same form in different S-frames.

It remains to prove the operator $ S $ defined through (\ref{abc1}) could make (\ref{u1}) satisfied. According to (\ref{u1}), for the supercharge $ Q $ and the special supercharge $ S $, 
\begin{eqnarray}\label{a1}
\nonumber&&  (\frac{c \tau +d}{c \bar{\tau} +d})^{-\frac{1}{4}}\hat{Q}^{a}_{\alpha}(\tau)=S^{-1}\hat{Q}^{a}_{\alpha}(\frac{a \tau+b}{c \tau+d} ) S\;,\;\;\;\;\;\;\;\;       (\frac{c \tau +d}{c \bar{\tau} +d})^{\frac{1}{4}}   \hat{\bar{Q}}^{a}_{\dot{\alpha}}(\tau)=S^{-1}\hat{\bar{Q}}^{a}_{\dot{\alpha}}(\frac{a \tau+b}{c \tau+d} )S\;,\\ \nonumber&& (\frac{c \tau +d}{c \bar{\tau} +d})^{\frac{1}{4}}\hat{S}^{a\dot{\alpha}}(\tau)=S^{-1} \hat{S}^{a\dot{\alpha}}(\frac{a \tau+b}{c \tau+d} )S \;,\;\;\;\;\;\;\;\;     (\frac{c \tau +d}{c \bar{\tau} +d})^{-\frac{1}{4}}     \hat{\bar{S}}^{a\alpha}(\tau)=S^{-1}\hat{\bar{S}}^{a\alpha}(\frac{a \tau+b}{c \tau+d} )S\;.\\
\end{eqnarray}
Based on the superconformal algebra, 
\begin{equation}
\{\hat{Q}^{a}_{\alpha},\hat{\bar{Q}}_{b\dot{\beta}}\}=2\sigma^{\mu}_{\alpha\dot{\beta}}\hat{P}_{\mu}\delta^{a}_{b}\;,\;\;\;\;\;\;\;\; \{\hat{S}_{a\alpha},\hat{\bar{S}}^{b}_{\dot{\beta}}\}=2\sigma^{\mu}_{\alpha\dot{\beta}}\hat{K}_{\mu} \delta^{b}_{a}\;,
\end{equation}
\begin{equation}
\{\hat{Q}^{a}_{\alpha},\hat{S}_{b\beta}\}= \epsilon_{\alpha\beta}(\delta^{a}_{b}\hat{D}+\hat{R}^{a}_{b})+\frac{1}{2}\delta^{a}_{b}\sigma^{\mu\nu}_{\alpha\beta}\hat{M}_{\mu\nu}\;,
\end{equation}
({\ref{a1}}) also gives 
\begin{equation}\label{wa}
 \hat{P}_{\mu}(\tau)=S^{-1}\hat{P}_{\mu}(\frac{a \tau+b}{c \tau+d} )S\;,\;\;\hat{M}^{\mu\nu}(\tau)=S^{-1}\hat{M}^{\mu\nu}(\frac{a \tau+b}{c \tau+d} )S\;,\;\;\hat{R}^{a}_{b}(\tau)=S^{-1}\hat{R}^{a}_{b}(\frac{a \tau+b}{c \tau+d} )S
\end{equation}
\begin{equation}
\hat{K}_{\mu}(\tau)=S^{-1}\hat{K}_{\mu}(\frac{a \tau+b}{c \tau+d} )S \;,\;\;\;\;\;\;\;\;
\hat{D}(\tau)=S^{-1}\hat{D}(\frac{a \tau+b}{c \tau+d} )  S\;,
\end{equation}
so it is sufficient to prove (\ref{a1}).

 Since $ \{W(C),T(C)\} $ compose the complete physical operators, it is enough to check the action of $Q,\bar{Q},S,\bar{S}  $ on $ W $ and $ T $ under the S transformation. In \cite{bps}, supersymmetries preserved by various BPS Wilson-'t Hooft operators are studied. It was shown that supercharges preserved by BPS Wilson loops and their magnetic counterparts are related by a four dimensional chiral transformation. This could serve as a proof for the S-duality invariance of the theory. Let $ \{A_{k}|k=1,2,\cdots,n\} $ represent a subset of supercharges and $ H_{ \{A_{k}|k=1,2,\cdots,n\} } $ denote the space of BPS Wilson operators annihilated by them. $S^{-1} H_{ \{A_{k}|k=1,2,\cdots,n\} } S$ will be the space of the dual BPS 't Hooft operators annihilated by $ \{S^{-1}A_{k}S|k=1,2,\cdots,n\} $. If $S^{-1} H_{ \{A_{k}|k=1,2,\cdots,n\} } S$ is also annihilated by $ \{B_{k}|k=1,2,\cdots,n\} $ which is related to $ \{A_{k}|k=1,2,\cdots,n\} $ by a chiral rotation, there will be $ \{S^{-1}A_{k}S|k=1,2,\cdots,n\} = \{B_{k}|k=1,2,\cdots,n\} $ since otherwise, $S^{-1} H_{ \{A_{k}|k=1,2,\cdots,n\} } S  $ would have the lower dimension. We arrive at the conclusion that the operator $ S $ making the Wilson operator transformed into the 't Hooft operator will also make the supercharges and the special supercharges transform with a $ U(1)_{Y} $ phase so the theory is duality invariant.

\subsection{S-duality transformation of gauge invariant operators}

In this section, we will study the S-duality transformation of gauge invariant operators and especially, the $ 1/2 $ BPS operators. 

\subsubsection{Generic}

In theory with the coupling constant $ \tau $, the generic on-shell gauge invariant operator $ O $ could be constructed from the canonical fields:
\begin{equation}
O(\tau)= O[X, \Psi,A;\Pi_{X},\Pi_{\Psi},\Pi_{A}|\tau]\;.
\end{equation}
The S-dual of $ O(\tau)$ is constructed from the dual canonical fields and the dual coupling constant: 
\begin{eqnarray}
\nonumber && \tilde{O}(\frac{a \tau+b}{c \tau+d})=  O[\tilde{X}, \tilde{\Psi},\tilde{A};\tilde{\Pi}_{X},\tilde{\Pi}_{\Psi},\tilde{\Pi}_{A}|\frac{a \tau+b}{c \tau+d}]\\ &=&
U^{-1}O(\frac{a \tau+b}{c \tau+d})
U=O'[X, \Psi,A;\Pi_{X},\Pi_{\Psi},\Pi_{A}|\tau]=
O'(\tau)\;.
\end{eqnarray} 
$O'(\tau)  $ is still an operator in theory with the coupling constant $ \tau $ and may be highly nonperturbative when written in terms of the original fields. The projection in $ H_{ph} $ is
\begin{equation}
\hat{\tilde{O}}(\frac{a \tau+b}{c \tau+d})=S^{-1}\hat{O}(\frac{a \tau+b}{c \tau+d})S= \hat{O}'(\tau)\;.
\end{equation}

The correlation functions of $ O'(\tau) $ and $O(\tau)  $ in theory with the coupling constant $ \tau $ are
\begin{eqnarray}\label{qa}
\nonumber &&	\langle 0| O'(\textbf{x}_{1},t_{1}|\tau) O'(\textbf{x}_{2},t_{2}|\tau) |0\rangle=\langle 0| O'(\textbf{x}_{1},0|\tau) e^{i\hat{H}[\tau](t_{2}-t_{1})}O'(\textbf{x}_{2},0|\tau) |0\rangle \\\nonumber &=&\langle 0'| O(\textbf{x}_{1},0|\frac{a \tau+b}{c \tau+d}) S(0)e^{i\hat{H}[\tau](t_{2}-t_{1})}S^{-1}(0)O(\textbf{x}_{2},0|\frac{a \tau+b}{c \tau+d}) |0'\rangle\\&=&\langle 0'| O(\textbf{x}_{1},0|\frac{a \tau+b}{c \tau+d}) e^{i\hat{H}[\frac{a \tau+b}{c \tau+d}](t_{2}-t_{1})}O(\textbf{x}_{2},0|\frac{a \tau+b}{c \tau+d}) |0'\rangle\;
\end{eqnarray}
and
\begin{equation}\label{qa1}
\langle 0| O(\textbf{x}_{1},t_{1}|\tau) O(\textbf{x}_{2},t_{2}|\tau) |0\rangle=\langle 0| O(\textbf{x}_{1},0|\tau) e^{i\hat{H}[\tau](t_{2}-t_{1})}O(\textbf{x}_{2},0|\tau) |0\rangle \;.
\end{equation}
The correlation function of $ O' $ in theory with the coupling constant $ \tau $ is equal to the correlation function of $ O $ in the same theory with the coupling constant $ \frac{a \tau+b}{c \tau+d} $ as is required in (\ref{cor}). To get (\ref{qa}), we used (\ref{harmi}), which, if is not satisfied, will make the correlation function of $ O' $ mapped into the correlation function of $ O $ in a different theory with the coupling constant $ \frac{a \tau+b}{c \tau+d} $.

In special case, if
\begin{equation}\label{1/2}
S^{-1}\hat{O}(x|\frac{a \tau+b}{c \tau+d})S=\hat{O}'(x|\tau) = (\frac{c \tau +d}{c \bar{\tau} +d})^{-\frac{q}{4}}\hat{O}(x|\tau)\;
\end{equation}
with $ q $ the $ U(1)_{Y} $ charge of $ O $, then according to (\ref{qa}), 
\begin{eqnarray}
\nonumber &&	\langle 0'| O(\textbf{x}_{1},0|\frac{a \tau+b}{c \tau+d}) e^{i\hat{H}[\frac{a \tau+b}{c \tau+d}](t_{2}-t_{1})}O(\textbf{x}_{2},0|\frac{a \tau+b}{c \tau+d}) |0'\rangle\\&=&(\frac{c \tau +d}{c \bar{\tau} +d})^{-\frac{q}{2}}\langle 0| O(\textbf{x}_{1},0|\tau) e^{i\hat{H}[\tau](t_{2}-t_{1})}O(\textbf{x}_{2},0|\tau) |0\rangle\;.
\end{eqnarray}
The correlation functions of $ O $ in theories with the coupling constants $ \frac{a \tau+b}{c \tau+d} $ and $ \tau$ differ by a $ U(1)_{Y} $ phase. Generically,
\begin{equation}\label{429}
\langle \prod_{i}O^{(q_{i})}_{i}(x_{i})\rangle_{\frac{a \tau+b}{c \tau+d}}=(\frac{c \tau +d}{c \bar{\tau} +d})^{-\frac{1}{4}\sum_{i}q_{i}}\langle \prod_{i}O^{(q_{i})}_{i}(x_{i})\rangle_{\tau}\;,
\end{equation}
where $ q_{i} $ is the $ U(1)_{Y} $ charge of $ O^{(q_{i})}_{i} $. This is the $ U(1)_{Y} $ transformation rule for the correlation function of $ 1/2 $ BPS operators conjectured in \cite{op1}.

\subsubsection{$ 1/2 $ BPS operators}

$ 1/2 $ BPS operators can be obtained by the successive action of the supercharges on chiral primary operators $  \hat{O}(\tau)=\tau_{2}^{k/2}\hat{tr}( X^{\{I_{1}}\cdots X^{I_{k}\}})  $. $ \hat{O}^{(m,n)}\sim \delta^{m}\bar{\delta}^{n}\hat{O} $ has the $ U(1)_{Y} $ charge $ q=m-n $. According to AdS/CFT, $ \hat{O}^{(m,n)} $ is mapped into the supergravity field $ F^{(m,n)} $. Under the $ SL(2,Z) $ transformation in type IIB string theory, 
\begin{equation}
F^{(m,n)}  \rightarrow (\frac{c \tau +d}{c \bar{\tau} +d})^{\frac{q}{4}} F^{(m,n)} \;.
\end{equation}
The low energy effective action of type IIB string theory is expected to be $ SL(2,Z) $ invariant and so the correlation functions of $ 1/2 $ BPS operators are conjectured to transform as (\ref{429}) \cite{op1}. Of course, if 
\begin{equation}\label{2/2}
\hat{O}'^{(m,n)}(x|\tau) = (\frac{c \tau +d}{c \bar{\tau} +d})^{-\frac{q}{4}}\hat{O}^{(m,n)}(x|\tau)\;, 
\end{equation}
(\ref{429}) can be automatically satisfied. Moreover, current multiplet is also $ 1/2 $ BPS, so (\ref{2/2}) would make the superconformal charges transform with a $ U(1)_{Y} $ phase thus could ensure the S-duality invariance of the theory. In the following, we will check the validity of (\ref{2/2}).

Suppose the gauge group is $ SU(2)$ and for simplicity, let $ \tau =i \tau_{2}$ and consider the transformation with $ \tau \rightarrow -1/\tau $. For
\begin{eqnarray}
 &&O^{[ab][cd]}(\tau)=\tau_{2} tr(2X^{ab}X^{cd}+X^{ac}X^{bd}-X^{ad}X^{bc})\\&& O^{a[cd]}_{\alpha}(\tau)=\tau_{2} tr(2\Psi^{a}_{\alpha}X^{cd}+\Psi^{c}_{\alpha}X^{ad}-\Psi^{d}_{\alpha}X^{ac})\\&&O^{(ab)}(\tau)=\tau_{2}  tr(-\Psi^{\alpha a}\Psi^{b}_{\alpha}+t^{(ab)}_{cdefgh}X^{cd}X^{ef}X^{gh})
\end{eqnarray}
with the $ U(1)_{Y} $ charges $ 0 $, $ 1 $ and $ 2 $, where $ O^{[ab][cd]} \sim O$ is the chiral primary operator, $ O^{a[cd]}_{\alpha} \sim \delta O$ and $ O^{(ab)} \sim \delta^{2}O$ are descendants, 
\begin{eqnarray}
 \label{q1}&& \hat{O}^{[ab][cd]}(\tau) \rightarrow \hat{O}^{[ab][cd]}(-1/\tau)\\ \label{q2}&& \hat{O}^{a[cd]}_{\alpha}(\tau) \rightarrow  e^{\frac{i \pi}{4}}\hat{O}^{a[cd]}_{\alpha}(-1/\tau)\\ \label{q3}&& \hat{O}^{(ab)}(\tau)\rightarrow  e^{\frac{i \pi}{2}}\hat{O}^{(ab)}(-1/\tau)
\end{eqnarray}
should be realized. For $ SU(2) $ group, at the classical level, (\ref{q1}) can only be possible if $ X^{ab} \rightarrow X^{ab}/\tau_{2}$ up to a gauge transformation. The multiplication of $ e^{\frac{i \pi}{4}}  $ for $  O^{a[cd]}_{\alpha}  $ requires $ \Psi^{a}_{\alpha}\rightarrow  e^{\frac{i \pi}{4}}\Psi^{a}_{\alpha} /\tau_{2}$. And then (\ref{q3}) cannot be satisfied unless the gauge group is $ U(1) $. The situation is the same with fields replaced by operators. The analysis does not rely on $ S $ just indicating $ O $, $ \delta O $ and $ \delta^{2}O $ cannot transform as in (\ref{q1})-(\ref{q3}) simultaneously. So, at least for some $ 1/2 $ BPS operators, (\ref{2/2}) is not valid.

The operator $ S $ should make (\ref{a1}) satisfied, so $ S $ cannot make the chiral primary operators $ \hat{O} $ remain invariant, otherwise, (\ref{2/2}) will always hold. In fact, the invariance of $ \hat{O} $ is a strong constraint which requires $ \tilde{X}^{I} =|c \tau +d|X^{I}$ in (\ref{u}) up to a gauge transformation.

It remains to determine how the operator $ S $ making Wilson operators transformed into the 't Hooft operators will act on $ \hat{O} $. With $S  \hat{O} S^{-1}$ given, $ X $ can be fixed up to a gauge transformation and then, for an arbitrary physical operator $ K(X) $ composed by $ X $, including all of the superconformal primary operators, $ S\hat{K}(X)S^{-1} $ is also determined. With (\ref{u1}) satisfied, the successive action of superconformal charges will then give the dual operators for the whole superconformal multiplet.

\subsection{S-duality transformation of physical states}

The physical Hilbert space $ H_{ph} $ forms a reducible representation of the superconformal group and could be decomposed into the direct sum of the irreducible subspaces: $H_{ph} =\oplus \;H^{(i)}_{ir}(\tau)  $. The decomposition is $ \tau $-dependent and the states in $ H_{ph} $ are required to be normalizable. The superconformal generators $ G (\tau)$ are then the block diagonal matrices in this representation: $G(\tau) =\oplus\; G^{(i)}_{ir} (\tau) $.

The global time Hamiltonian is $ h=\frac{1}{2}(K_{0}+P_{0}) =\frac{1}{2}(K_{0}+H)$,  
\begin{equation}
 S\hat{h}(\tau)S^{-1}=\hat{h}(\frac{a\tau+b}{c\tau+d})\;.
\end{equation}
$\hat{h}  $ has the normalizable eigenstates $ \{|E_{i}\rangle\} $ with the discrete eigenvalues $ \{E_{i}\} $,
 \begin{equation}
\hat{h} (\tau)|E_{i}(\tau)\rangle=E_{i}(\tau)|E_{i}(\tau)\rangle\;.
 \end{equation}
Since
  \begin{equation}
\hat{h} (\frac{a\tau+b}{c\tau+d})|E_{i}(\frac{a\tau+b}{c\tau+d})\rangle=E_{i}(\frac{a\tau+b}{c\tau+d})|E_{i}(\frac{a\tau+b}{c\tau+d})\rangle\;,
 \end{equation}
 we have
\begin{equation}
\hat{h}(\tau)S^{-1}|E_{i}(\frac{a\tau+b}{c\tau+d})\rangle=E_{i}(\frac{a\tau+b}{c\tau+d})S^{-1}|E_{i}(\frac{a\tau+b}{c\tau+d})\rangle\;.
 \end{equation}
If $ |E_{i}(\tau)\rangle $ is a normalizable eigenstate of $\hat{h}(\tau)  $ with the eigenvalue $ E_{i}(\tau) $, $   S^{-1}|E_{i}(\frac{a\tau+b}{c\tau+d})\rangle$ will be a normalizable eigenstate of $\hat{h}(\tau)  $ with the eigenvalue $ E_{i}(\frac{a\tau+b}{c\tau+d}) $. The spectrum of $  \hat{h}(\tau)$ is $ SL(2,Z) $ invariant: $ \{E_{i}(\tau)\}= \{E_{i}(\frac{a\tau+b}{c\tau+d})\}$.

Generically, for a state $|f(\tau)\rangle  $, the dual state is $S^{-1} |f(\frac{a\tau+b}{c\tau+d})\rangle  $. In each irreducible representation, if $|f_{0}(\tau)\rangle  $ is the superconformal primary state with the energy $ E_{0} (\tau)$, $G(\tau)\cdots G(\tau)|f_{0}(\tau)\rangle  $ will give the whole multiplet. The S-dual of $|f_{0}(\tau)\rangle  $ is the superconformal primary state $S^{-1} |f_{0}(\frac{a\tau+b}{c\tau+d})\rangle  $ with the energy $ E_{0} (\frac{a\tau+b}{c\tau+d})$, while the S-dual of $G(\tau)\cdots G(\tau)|f_{0}(\tau)\rangle  $ is $S^{-1}G(\frac{a\tau+b}{c\tau+d})\cdots G(\frac{a\tau+b}{c\tau+d})|f_{0}(\frac{a\tau+b}{c\tau+d})\rangle  \sim   G(\tau)\cdots G(\tau)S^{-1} |f_{0}(\frac{a\tau+b}{c\tau+d})\rangle$. So the duality maps one irreducible representation into another. State and the dual state are in the same $ SO(3)\times SU(4) $ representation with the energies $ E(\tau) $ and $ E(\frac{a\tau+b}{c\tau+d}) $, respectively.

 A normalizable state $ |O(\tau)\rangle $ in $ H_{ph} $ corresponds to a renormalized operator $ O_{r} (\tau)$ with the finite two point function \cite{8}.
 \begin{equation}
  |O(\tau)\rangle=e^{\frac{\pi}{4}(H-K_{0})} O_{r} (0|\tau)|0\rangle\;.
 \end{equation}
 $O_{r}   $ can be expanded in terms of the bases $\{O^{I}_{r}   \}$ with $ O^{I}_{r} (\tau) =Z^{I}_{J} (\tau)O^{J} (\tau)$. $O^{J} (\tau)  $ is the bare operator and $ Z^{I}_{J} (\tau) $ is the renormalization matrix. The $ \tau $-dependence of $O_{r} (\tau)  $ also comes from the renormalization. If 
  \begin{equation}
\hat{h} (\tau)|O(\tau)\rangle=\Delta(\tau)|O(\tau)\rangle\;, 
 \end{equation}
 then
 \begin{equation}
[\hat{D}(\tau), \hat{O}_{r} (\tau)]=i\Delta(\tau) \hat{O}_{r} (\tau)\;. 
\end{equation}
 The dual operator of $ \hat{O}_{r} (\tau) $ is $ \hat{O}'_{r} (\tau)=S^{-1}\hat{O}_{r} (\frac{a\tau+b}{c\tau+d}) S$ related with the normalizable state $ S^{-1}|O(\frac{a\tau+b}{c\tau+d})\rangle $ and so, are still properly renormalized with the finite two point function. 
   \begin{equation}
[\hat{D}(\frac{a\tau+b}{c\tau+d}), \hat{O}_{r} (\frac{a\tau+b}{c\tau+d})]=i\Delta(\frac{a\tau+b}{c\tau+d}) \hat{O}_{r} (\frac{a\tau+b}{c\tau+d})\;.
\end{equation}
Since $  S\hat{D}(\tau)S^{-1}=\hat{D} (\frac{a\tau+b}{c\tau+d})$, 
 \begin{equation}
[\hat{D}(\tau), \hat{O}'_{r} (\tau)]=i\Delta(\frac{a\tau+b}{c\tau+d}) \hat{O}'_{r} (\tau)\;. 
\end{equation}
 The spectrum of conformal dimension, which is the same as the spectrum of $\hat{h}  $, is $ SL(2,Z) $ invariant: $ \{\Delta_{i}(\tau)\}= \{\Delta_{i}(\frac{a\tau+b}{c\tau+d})\}$.

 Now consider the chiral primary operator $\hat{O}(\tau)=\tau_{2}^{k/2}\hat{tr}( X^{\{I_{1}}\cdots X^{I_{k}\}})   $ and the corresponding state $ |O(\tau)\rangle $, $ \hat{O}'(\tau) \neq \hat{O}(\tau)  $ according to the previous discussion. However, as the $ SO(3) $ invariant state with the energy $ k $ in $ (0,k,0) $ representation of $ SU(4) $, $ |O(\tau)\rangle $ is unique, so $ |O(\tau)\rangle= |O'(\tau)\rangle  $ based on (\ref{u1}). We have to assume
 \begin{equation}
[ \hat{O}'(\tau) - \hat{O}(\tau)]|0\rangle=0\;.
 \end{equation}
The difference between $ \hat{O}' $ and $ \hat{O} $ annihilates the vacuum. Successive action of the supercharges gives
 \begin{equation}\label{lab}
[ \hat{O}'^{(m,n)}(\tau) -(\frac{c \tau +d}{c \bar{\tau} +d})^{-\frac{q}{4}} \hat{O}^{(m,n)}(\tau) ]|0\rangle=0\;.
 \end{equation}
With (\ref{lab}) satisfied, correlation functions of $ 1/2 $ BPS operators at different points could respect the $ U(1)_{Y} $ transformation rule. In fact, it is also emphasized in \cite{op3} that $ U(1)_{Y} $ rule only applies for the separated operators.

The coincident $ 1/2 $ BPS operators could compose the non-BPS unprotected multi-trace operators related with the bound states in AdS. For example, $ \hat{O}(\tau) \hat{O}(\tau)|0\rangle$ and $\hat{O}'(\tau)\hat{O}'(\tau)|0\rangle  $, when properly renormalized, will correspond to the two particle bound states with the coupling constant dependent energy. State and its S-dual have the different energies thus could not be the same any more. The difference between $ \hat{O}' $ and $ \hat{O} $ may have the manifestation here.

\subsection{$ \theta $ structure and the large gauge transformation}

It is necessary to consider the impact of the S-duality transformation on the global structure of the gauge theory. In temporal gauge, time-independent gauge transformation is realized through the unitary operator $ \Xi(\omega(x^{i})) $ generating a $ U(N) $ gauge transformation $ u(\omega(x^{i}))=e^{i  t^{A}\omega_{A}} $. When $ |\vec{x}|\rightarrow \infty $, $  u(\omega(x^{i}))\rightarrow I$. $ \Pi_{3}(U(N))\cong Z $, $ u(\omega) $ is classified by its winding number $ n $,
\begin{equation}
n=\frac{1}{24 \pi^{2}}\int d^{3}x\; \epsilon_{ijk}tr  [(u^{-1}\partial_{i}u )(u^{-1}\partial_{j}u )(u^{-1}\partial_{k}u) ]\;.
\end{equation}
$ \Xi$ and $ u$ with the winding number $ n $ are denoted as $ \Xi^{(n)}$ and $ u^{(n)} $. $ \Xi^{(n)}\Xi^{(m)}=\Xi^{(n+m)} $, $ u^{(n)}u^{(m)}=u^{(n+m)} $ \cite{qcd}.

$ H_{ph} $ could be decomposed into the ``direct integral" of the subspaces $ H_{ph\;\theta} $. $ \theta \cong \theta + 2\pi  $. $ \forall\;\vert \psi\rangle_{\theta} \in H_{ph\;\theta} $, \cite{jack}
\begin{equation}
\Xi^{(n)} \vert \psi\rangle_{\theta}=e^{in\theta}\vert \psi\rangle_{\theta}\;.
\end{equation}
Especially, $\Xi^{(0)} \vert \psi\rangle=\vert \psi\rangle $, $ \forall\;\vert \psi\rangle \in H_{ph} $. For a theory with the coupling constant $ \tau = \frac{\theta}{2 \pi} +i \frac{4 \pi}{g^{2}}$, the true physical Hilbert space is $ H_{ph\;\theta} $. Suppose $ \vert 0\rangle_{\theta} $ is the vacuum in $H_{ph\;\theta}  $, states in $ H_{ph\;\theta}   $ can be constructed as $ \hat{O} \vert 0\rangle_{\theta} $ with $\hat{O}    $ the gauge invariant operator, $ [\Xi^{(n)}, \hat{O} ]=0 $, $ \forall\; n $.

Under the S-duality transformation, $ \tau\rightarrow \tau' =(a \tau+b)/(c \tau+d)= \frac{\theta'}{2 \pi} +i \frac{4 \pi}{g'^{2}}$, we may have
\begin{equation}
S[ \left(                
\begin{array}{cc}   
a & b\\ 
c & d \\  
\end{array}
\right) ;\tau]\vert 0\rangle_{\theta}  =\vert 0\rangle_{\theta'}  \;.
\end{equation}
Accordingly, in theory with the coupling constant $ \tau $, the S-dual of $\vert 0\rangle_{\theta}     $ should be 
\begin{equation}
\vert 0\rangle_{\theta}'   =S^{-1}\vert 0\rangle_{\theta'} =  \vert 0\rangle_{\theta} 
\end{equation}
which is the same as $ \vert 0\rangle_{\theta} $ as is required.

Generically, $ \forall\; \vert \psi(\tau)\rangle_{\theta} \in  H_{ph\;\theta}$, the dual state is $  \vert \psi(\tau)\rangle'_{\theta}=S^{-1}\vert \psi(\tau')\rangle_{\theta'} \in H_{ph\;\theta}$. Since 
\begin{equation}
\Xi^{(n)} \vert \psi(\tau)\rangle_{\theta} =e^{in\theta}\vert \psi(\tau)\rangle_{\theta}  \;\;\;\;\;\;\;\;\;\; \Xi^{(n)} \vert \psi(\tau')\rangle_{\theta'} =e^{in\theta'}\vert \psi(\tau')\rangle_{\theta'} \;,
\end{equation}
there will be  
\begin{equation}
S^{-1}\Xi^{(n)} S \vert \psi(\tau)\rangle'_{\theta} =e^{in\theta'}\vert \psi(\tau)\rangle'_{\theta}=
e^{in(\theta'-\theta)}\Xi^{(n)}   \vert \psi(\tau)\rangle'_{\theta} \;.
\end{equation}

We may expect
\begin{equation}
S^{-1}[ \left(                
\begin{array}{cc}   
a & b\\ 
c & d \\  
\end{array}
\right) ;\tau]\Xi^{(n)} S[ \left(                
\begin{array}{cc}   
a & b\\ 
c & d \\  
\end{array}
\right) ;\tau]=e^{in(\theta'-\theta)}\Xi^{(n)} \;,
\end{equation}
or equivalently, 
\begin{equation}
\Xi^{(n)} S[ \left(                
\begin{array}{cc}   
a & b\\ 
c & d \\  
\end{array}
\right) ;\tau]\Xi^{(n)-1} =e^{in(\theta'-\theta)}S[ \left(                
\begin{array}{cc}   
a & b\\ 
c & d \\  
\end{array}
\right) ;\tau]\;. 
\end{equation}
Although $ S $ commutes with $ \Xi^{(0)} $, it does not necessarily commute with $ \Xi^{(n)} $ for $ n\geq 1 $.

As the simplest example, consider the transformation $ \tau \rightarrow \tau+\frac{\phi}{2 \pi} $ with
\begin{equation}
U(\phi)\Pi^{A}_{i} U^{-1}(\phi)= \Pi^{A}_{i} +\frac{\phi}{2 \pi}  B^{A}_{i} \;\;\;\;\;\;\;\;U(\phi)A^{A}_{i} U^{-1}(\phi)= A^{A}_{i} \;.
\end{equation}
The unitary operator $ U (\phi)$ realizing such transformation is explicitly constructed \cite{cons}:
\begin{equation}\label{top}
U(\phi)=e^{i \phi q}\;,
\end{equation}
where 
\begin{equation}
q=\int d^{3}x\; K^{0}(x)
\end{equation}
is the topological charge. $K^{0}  $ is the time component of the current
\begin{equation}
K^{\mu}=\frac{1}{8\pi}\epsilon^{\mu\nu\lambda\sigma}(A^{A}_{\nu}F^{A}_{\lambda\sigma}-\frac{1}{3}f^{ABC}A_{\nu}^{A}A_{\lambda}^{B}A_{\sigma}^{C})\;.
\end{equation}
For the $ \theta $ vacuum 
\begin{equation}
\vert 0 \rangle_{\theta} \sim   \sum_{n}  e^{i n \theta}\vert n \rangle
\end{equation}
with $ \vert n \rangle $ representing the pure gauge state with the winding number $ n $, 
\begin{equation}
U(\phi)\vert 0 \rangle_{\theta}=\vert 0 \rangle_{\theta+\phi}\;.
\end{equation}

Since
\begin{equation}
\Xi^{(n)}  q \;\Xi^{(n)-1}=q+n\;,\;\;\;\;\;\;\;\;\;\;
\Xi^{(n)}  e^{i\phi q} \;\Xi^{(n)-1}= e^{i\phi n} e^{i\phi q}\;,
\end{equation}
for 
\begin{equation}
S[ \left(                
\begin{array}{cc}   
1 & \frac{\phi}{2 \pi}\\ 
0& 1 \\  
\end{array}
\right) ;\tau]=e^{i\phi q}\;,
\end{equation}
there will be 
\begin{equation}
\Xi^{(n)}  S[ \left(                
\begin{array}{cc}   
1 & \frac{\phi}{2 \pi}\\ 
0& 1 \\  
\end{array}
\right) ;\tau] \;\Xi^{(n)-1}= e^{i\phi n} S[ \left(                
\begin{array}{cc}   
1 & \frac{\phi}{2 \pi}\\ 
0& 1 \\  
\end{array}
\right) ;\tau]\;.
\end{equation}

\section{Discussion}

S-duality transformation in loop space can be explicitly realized as an $ SL(2,Z) $ canonical transformation. However, there is a gap between the loop space and the ordinary space, and the question is how the same $ S $ will act on the local gauge invariant operators. Since local operators also appear in the OPE of loop operators \cite{11, 12, wh4}:
\begin{equation}\label{wls}
W(^{L}R)=\langle W(^{L}R) \rangle (1+\sum_{i}  c_{i} a^{\Delta_{i}} O_{i})\;,
\end{equation}
\begin{equation}
T(^{L}R)=\langle T(^{L}R) \rangle (1+\sum_{i} b'_{i} a^{\Delta'_{i}}O'_{i})\;,
\end{equation}
by studying the OPE of Wilson and 't Hooft operators, the S-dual local operators can be obtained. For Wilson operator, the right hand side of (\ref{wls}) can be written explicitly \cite{12}, but the challenge is to get $O'_{i}  $ from $ T(^{L}R) $ which has a quantum mechanical definition in path integral formalism \cite{wh3}.

The commutation relation between the Wilson and 't Hooft operators leads to a natural canonical commutation relation in loop space with the linking number $ n(C,C') $ playing the role of $ \delta^{3} (\vec{x}-\vec{x}') $. It seems that S-duality transformation rule always takes a more simple form for loop operators. It is necessary to study the loop space formulation of $ \mathcal{N} $ $ =4 $ SYM theory in more detail, especially, with the super Wilson operators \cite{su1, su2, su3}.

The operator $ S $ is defined through its action on operators which are both local and global gauge invariant. The vacuum is also selected to be global gauge invariant, so our discussion is actually restricted to the global gauge invariant Hilbert space $ H_{g, ph}$ which is a subspace of $ H_{ph} $, $ H_{g, ph} \subset  H_{ph} $. This is consistent with AdS/CFT, where the bulk theory is dual to the colorless sector of the gauge theory. However, on gauge theory side, states in $ H_{ph} $ only need to be local gauge invariant and could form the representation of the global gauge transformation. The global gauge transformation operator is physical, and the S-dual of which should also be constructed. In conformal phase, we do not have too many clues to define it. Maybe it is better to address the problem in Coulomb phase, where the electric charge and the magnetic charge can be obtained as the central charge of the superalgebra \cite{122}.

Type IIB string theory in $ AdS_{5} \times S^{5}$ is dual to the $ \mathcal{N} $ $ =4 $ SYM theory. With the type IIB Hilbert space in one-to-one correspondence with the SYM physical Hilbert space, $ \mathcal{N} $ $ =4 $ SYM theory gives a definition for the quantum Type IIB theory. We may expect the S-duality transformation in Type IIB theory can also be realized via an operator $ S $, under which, the $ PSU(2,2|4) $ charges transform with the $ U(1)_{Y} $ phase and the AdS fields $ f^{(1,0)}_{m,n} $ related with the $ (1,0) $ string are mapped into the fields $ f^{(a,b)}_{m,n} $ for $ (a,b) $ string.

\bigskip
\bigskip

\section*{Acknowledgments}

The author would like to thank Ken Intriligator, I\~{n}aki Garc\'{i}a-Etxebarria and Hai Lin for helpful comments and discussions. The work is supported in part by NSFC under the Grant No. 11605049.


\bibliographystyle{plain}

\end{document}